# Evidence of room temperature magnetoelectric properties in $\alpha$-Fe$_{1.6}$Ga$_{0.4}$O$_3$ oxide by magnetic field controlled electric properties and electric field controlled magnetism

R.N. Bhowmik[*], and Abdul Gaffar Lone

Department of Physics, Pondicherry University, R.V. Nagar, Kalapet, Pondicherry-605014, India

[*]Corresponding author: Tel.: +91-9944064547; Fax: +91-413-2655734

E-mail: rnbhowmik.phy@pondiuni.edu.in

**Abstract**

We have stabilized the $\alpha$-Fe$_{1.6}$Ga$_{0.4}$O$_3$ (Ga doped $\alpha$-Fe$_2$O$_3$) system in rhombohedral structure. The system has shown magnetically canted ferromagnetic state and ferroelectric properties at room temperature. In first time, we confirm the existence of magneto-electric coupling and multiferro-electric properties at room temperature in the Ga doped $\alpha$-Fe$_2$O$_3$ system based on the experimental observation of magnetic field controlled electric polarization and electric field controlled magnetization. The $\alpha$-Fe$_2$O$_3$ system does not exhibit electric field controlled magnetic exchange bias effect, where as Ga doped $\alpha$-Fe$_2$O$_3$ showed exchange bias shift up to the value of 370 Oe. We have recorded the response of current-voltage characteristics and in-field magnetic relaxation of the system under the simultaneous application of magnetic and electric fields. The magnetization of the system is found highly sensitive to the ON and OFF modes, as well as change of the polarity of external electric field. The system is a new addition in the list of non-traditional magneto-electrics/multi-ferroelectrics so far reported in literature. Such novel materials, where magnetization and electric polarization can be controlled by simultaneous application of magnetic and electric fields, is in the increasing demand for potential applications in the field of next generation magnetic sensor, switching, non-volatile memory and spintronic devices.

# 1. Introduction

Two independent land mark observations, (1) magnetism in a moving dielectric under electric field and (2) electric polarization in a moving dielectric under magnetic field, laid the founding stone of the concept of magneto-electric effect [1]. Antiferromagnetic $Cr_2O_3$ was the first single phased system that provided experimental evidence of the linear magneto-electric effect, i.e., electric current/field-induced magnetization [2] and magnetic field induced polarization [3]. However, magneto-electric coupling in $Cr_2O_3$ oxide is not strong enough to make the material suitable for room temperature applications. This intensified research interest for searching single phase new materials having sufficiently strong coupling between magnetic spin order (magnetization) and electric charge order (polarization) at room temperature. The electric field controlled magnetic properties has been observed in different types of materials, including ferromagnetic metals, ferromagnetic semiconductor and multiferroics [4]. In case of ferromagnetic metals, one cannot apply large electric field due to screening effect. In case of ferromagnetic semiconductors, electric field alters electrical conductivity by changing the carrier concentration and mobility. The carrier concentration in a bulk ferromagnetic semiconductor is low enough to achieve strong electric field controlled magnetism, although the ferromagnetic semiconductor film of (In,Mn)As showed electric field controlled magnetic state, of course at temperatures well below of room temperature, for the first time [5]. The conventional spintronics devices, e.g., magnetic tunnel junctions (MTJs) and the devices using spin Hall effect, work based on spin-transfer torque (STT) technique [4]. Such technique uses spin-polarized current or magnetic field to manipulate magnetism/switching of magnetization. The main problem of the current or magnetic field controlled magnetization switching is the Joule heating effect. The overheating effect can be minimized by electric field controlled magnetization switching in multiferroelectrics, where one needs to charge and discharge a capacitor and whose operation scheme is compatible with existing semiconductor technology [6].

Multiferroelectric materials belong to a typical class of magneto-electrics, where two of the ferroic orders (ferromagnetization, ferroelectric polarization and ferroelasticity) exist in the single phase system. In multiferroics, a strong coupling between electrical polarization and magnetic spin order allows the magneto-electric properties to be mutually controlled by electric and magnetic fields. The materials having room temperature multiferroelectric properties (coexistence of ferromagnetism and ferroelectricity, and a strong mutual coupling) can be used for a variety of applications in the field of spintronics, non-volatile memory devices, magnetic data storage devices (electric-writing and magnetic-reading), switching and sensor applications [6-9]. The ideal multi-ferroelectrics, where spontaneous polarization and magnetization coexist in same crystal structure, are very rare in nature due to mutually exclusive nature of ferroelectric (needs empty d shell ions) and magnetic (needs partially filled d shell ions) orders. From symmetry consideration in crystal structure [7], ferroelectricity needs the breaking of spatial inversion symmetry while the time reversal symmetry can be invariant. In contrast, magnetic spin order needs the breaking of time-reversal symmetry while spatial-inversion symmetry can be invariant. Some oxides ($BiFeO_3$, $TbMnO_3$, and $CoCr_2O_4$), which contained magnetic elements in crystal structure, indicated multi-ferroelectric properties [10-12]. Despite the effort of searching new multiferroics and magnetoelectrics, the number of single-phased multiferroics is limited due to weak coupling between the magnetic unit (either antiferromagnetism or weak ferromagnetism) and electric unit at room temperature. Till date, only few materials have provided a direct or indirect evidence of the magneto-electric coupling effect where electric polarization can be controlled by magnetic field or magnetization can be controlled by electric field [13]. The electric field controlled switching of magnetization vectors can demonstrate the energy-efficient control of spin-valve device, where the required energy per unit area is less than that needed for spin-transfer torque switching process. Subsequently, the technique of electric field controlled magnetization switching is preferable for engineering magnetoelectric switching and development of low power consumption spintronic devices with additional non-volatile functionality by electric-field controlled magnetism [4].

On the other hand, searching of new magneto-electrics in transition metal oxides with room temperature ferromagnetic and ferroelectric properties has emerged as a new field of research [14]. The hematite (α-$Fe_2O_3$) is electrically an anisotropic insulator and does not show ferroelectric properties. But, hematite derived systems ($GaFeO_3$, $FeTiO_3$) and a series of metal doped oxides ($MnTiO_3$, $NiTiO_3$, $ScFeO_3$) offered magneto-electric properties over a wide temperature range [15-22]. Recently, Ga doped α-$Fe_2O_3$ system exhibited enhancement of ferromagnetic moment, tailoring of optical band gap in the range 2.2-3.0 eV, low dielectric loss, reasonably good dielectric constant, and signature of ferroelectric and magneto-electric properties [23-29]. The Ga doped hematite system seems to be a promising magnetoelectric material at room temperature [30]. In this work, we provide experimental evidences that α-$Fe_{1.6}Ga_{0.4}O_3$ could be a new magneto-electric and a possible multi-ferroelectric system, where magnetization and electric polarization can be controlled by the simultaneous application of magnetic and electric fields at room temperature.

## 2. Experimental

We have reported the details of material preparation and characterization elsewhere [26-27]. The compound α-$Fe_{1.6}Ga_{0.4}O_3$ was prepared by mechanical alloying of the fine powders of α-$Fe_2O_3$ and β-$Ga_2O_3$ oxides with alloying time up to 100 hrs. The alloyed powder was pressed into pellets (∅~13mm, t ~1mm), and annealed at 800 °C. The heating and cooling process was performed under high vacuum ($10^{-6}$ mbar) to avoid the formation of orthorhombic phase. X-ray diffraction pattern and Raman spectra confirmed single phase rhombohedral structure with space group $R\bar{3}C$. In this work, we have selected two samples MA50V8 (alloying time 50 h, grain size ~37 nm, lattice parameters: a = 5.0369Å, c = 13.744Å, V= 301.99Å$^3$) and MA100V8 (alloying time 100 h, grain size ~26nm, a = 5.0352Å, c = 13.715Å, V= 301.15Å$^3$) for showing magneto-electric properties in Ga doped hematite system. The pellet shaped samples were sandwiched between two Pt plates (electrodes) using a homemade sample holder. The electrical connection from Pt electrodes to Keithley Source Meter (2410-C) was made using shielded Pt wires for the measurement of current vs. voltage characteristics. The Pt wires were taken out from both sides of the sample and connected to Precision Premier II ferroelectric loop tester (Radiant Tech., USA) for electric polarization vs. electric field measurement. The sample was heated up to 160 $^0$C and subsequently, cooled down to room temperature in the presence of dc electric field 1 kV (i.e., the sample was electrically field cooled or poled) and P-E loop was recorded at room temperature. The electrical measurements under external magnetic field were performed by placing the sample holder at the pole gap of an electromagnet (MicroSense, USA). The magnetic field dependence of dc magnetization (M(H)) of the samples was measured using vibrating sample magnetometer (Lake Shore 7404, USA). A small pellet shaped sample of typical dimension 3mmx2mmx1 mm was placed between two thin Pt sheets, which were connected to 2410-C meter using thin Pt wires for applying dc electric voltage during the measurement of dc magnetization with magnetic field and time. The sample sandwiched between Pt electrodes was placed on the flat surface of the VSM sample holder (Kel-F) and tightly fixed with Teflon tape. The proper electrical contact has been checked from identical values of current on reversing the applied voltage at ±5 V.

### 3. Experimental Results

We show the typical P-E loops for MA50V8 sample, measured with the variation of voltage up to 1.5 kV at frequency 100 Hz (Fig. 1(a)), loop frequency down to 3.3 Hz at electric voltage of 1 kV (Fig. 1(b)), and magnetic field by sweeping the voltage up to 1 kV at 100 Hz (Fig. 1(c)), respectively. The P-E loops have indicated a good signature of switchable electric polarization with respect to the variation of electric field (voltage), loop frequency, and magnetic field. It is worthy to mention that the electric poling, while the sample was cooled @1 kV from 160 $^0$C to room temperature, improved the magnitude of electric polarization more than 10 times in comparison to the values measured without thermally cooled poling process [26]. The present values of polarization in Ga doped hematite system in rhombohedral phase are comparable to the polarization recently reported for Mg doped $Ga_{0.6}Fe_{1.4}O_3$ thin film in orthorhombic phase [31]. We emphasized on four important points (A, B, C, D as indicated in Fig. 1(a)) of the P-E curve to understand the effect of applied parameters (voltage, frequency and magnetic field). The point A represents the electric coercivity ($E_C$), where electric polarization is zero before reversal of direction. The point B represents the remanent polarization ($P_R$) when electric field becomes zero before reversal of direction. The point C represents the locus on P-E curve ($E_{max}$, $P_{Emax}$) at the value of maximum applied electric field ($E_{max}$). The general trend is that polarization (also remanent polarization ($P_R$)) and electric coercivity ($E_C$) of the material have increased with the decrease of frequency, and with the increase of electric field and magnetic field. A minor exception is noted for the variation of $P_R$ and $E_C$, where $P_R$ and $E_C$ showed an increasing trend above the loop frequency 50 Hz. On the other hand, a peak like feature appears in the P-E curve on decreasing the frequency, and increasing the magnitude of electric voltage and magnetic field. Such local polarization maximum ($P_{max}$) at the electric field $E_m$ has been represented by the point at D ($E_m$, $P_{max}$). The polarization at point C has considerably reduced than the values at point D ($P_{max} \geq P_{Emax}$). This shows an increase of leakage of polarization at extremely lower frequency (< 50 Hz) and higher magnetic field. One cannot completely neglect the conductive effect in semiconductor material, although capacitor effect dominates at room temperature [26]. The loss of polarization (LP(%)) due to conductive effect at electric fields above $E_m$ can be defined as LP (%) $=\frac{(P_{max}-P_{Emax})*100}{P_{max}}$ and the retaining of polarization (RP (%)) at zero electric field due to capacitive effect can be defined as RP (%) $= \frac{P_R*100}{P_{max}}$. In the present sample, LP showed minimum value (~4.3 %) at 50 Hz and it increases to 14.5 % at 3.3 Hz for electric voltage 1 kV. In the presence of magnetic field, the LP increases from 4.6 % at 1 kOe field to 12.3 % at 15 kOe. These values of the leakage of polarization are extremely low in comparison to 100 % as expected for a purely conductive effect in P-E loop [32]. A substantial increase of LP(%) on lowering the frequency may be affected by the space charge polarization effect at the interface of material and Pt electrodes, and the increase of polarization leakage with magnetic field may be associated with magneto-conductivity effect in the material [33]. On the other hand, retaining of polarization (RP) significantly increased from 28 % at 50 Hz to 58 % at 3.3 Hz of 1 kV. The retaining of polarization increased from 35 % at 1 kOe field to 51 % at 15 kOe. Similarly, retaining of polarization has significantly improved with the increase of applied voltage at the loop frequency 100 Hz (17 % at 500 V and 43 % at 1.5 kV) in comparison to the corresponding increase of leakage of polarization (2 % at 500 V and 8 % at 1.5 kV). Hence, the overall increase of polarization and retaining of polarization in the material cannot be treated absolutely due to conductivity effects.

We performed two independent experiments, i.e., magnetic field effect on electrical conductivity and electric field effect on magnetism, to confirm the magneto-electric coupling in the material. We measured current-voltage (I-V) curve with bias voltage in the range 0-500 V. Fig. 2 (a) shows a typical non-linear I-V characteristics of the MA100V sample. The I-V curve is also measured under 15 kOe magnetic field at ON (vertical dotted line) and OFF (vertical solid line) modes with 50 V intervals, i.e., magnetic field OFF at 0-50 V and magnetic field 15 kOe ON at 50 V-100 V. Such sequence of applying the magnetic field in OFF and ON mode was repeated while bias voltage swept from 0 V to 500 V. The I-V curve in the absence of magnetic field is smooth without any anomaly. The response of the I-V curve to the ON and OFF modes of magnetic field is reproducible. The change of electrical current at the point of magnetic field ON is not clearly distinguishable from the as usual increasing trend of current, whereas a sharp decrease of current is observed at the point of magnetic field OFF. First order derivative of the I-V curve (Fig. 2(b)) confirmed no abrupt change in the increasing trend of conductivity at the point of magnetic field ON unlike a spike that represents decrease of conductivity of the sample at the point of magnetic field OFF. The depth of spike increases on increasing the electric bias voltage. It shows that magneto-electric coupling; rather than magneto-conductivity effect, determines the magnetic field controlled I-V curve in the sample and magneto-electric coupling is expected to be higher at higher electric field [1]. We conclusively provide the evidence of room temperature magnetoelectric coupling and a possible multiferroelectric properties in the samples using the experimental results of electric field controlled magnetic state. First, we show in Fig. 3(a) the M(H) loop under different values of applied electric voltage (0-300 V) across the hematite ($\alpha$-$Fe_2O_3$) sample. The inset of Fig. 3(a) (magnified loop at 0 V and 300 V) confirms the absence of electric field induced shift of M(H) loop or exchange bias effect in hematite sample. The in-field magnetic relaxation of hematite sample (Fig. 3(b)) shows a typical character of an antiferromagnet or canted antiferromagnet [34]. The new observation is that magnetization vector of hematite sample, irrespective of the increasing or decreasing trend of magnetization with time (Fig. 3(c-d)), is highly sensitive to polarity, as well as to switching ON and OFF modes of the external electric voltage. The magnetization jumps instantly to higher state (lower state) by switching ON the electric voltage +100 V (-100 V) and returned back to original magnetic state by switching OFF the electric voltage. The response of magnetization with electric voltage at switched ON and OFF modes is repeatable and the change of magnetization is marked by nearly 0.87-1.00 %. The electric field controlled magnetic properties in Ga doped hematite system (Fig. 4 and Fig. 5) are remarkably different in comparison to hematite sample. Fig. 4(a) shows the M(H) loop of the MA100V sample, measured in the magnetic field range ±16 kOe and in the presence of constant electric voltage 0 V to + 200 V. The M(H) loops under positive electric voltage shifted towards the positive magnetic field axis and negative magnetization direction in comparison to the M(H) loop measured at 0 V. The shift of M(H) loop under electric voltage is clearly visible from the plot within small magnetic field range (Fig. 4(b)). The magnetic exchange bias field, $H_{exb}$, has been calculated from the shift of the centre of the loop under electric voltage with respect to the centre of the loop under 0 V. Fig. 4(c) shows that shift of the M(H) loop depends on polarity of the electric voltage. In contrast to the $180^0$ reversal of magnetization vector upon reversing the polarity of magnetic field, M(H) loop shifted towards positive magnetic field direction upon application of + 200 V (positive) and the loop shift under negative voltage (-200 V) is not exactly the mirror image of the loop shift measured under positive voltage (+ 200 V) with reference to the loop at 0 V. On reversing the electric field from + 200 V to -200 V, the magnetization shifted towards the negative magnetic field and positive magnetization directions when measured at -200 V in comparison to the M(H) loop measured at + 200 V. However, both the M(H) loops under opposite polarity of electric voltage remained in the positive side of the magnetic field with respect to the loop at 0 V. This results in a decrease of positive magnetic exchange bias field by nearly -75 Oe during measurement at - 200 V (with $H_{exb}$ = +267 Oe) with reference to the exchange bias field ($H_{exb}$ = +392 Oe) during measurement at + 200 V. The results are similar to that observed in the synthetic multiferroelectric film of $La_{0.67}Sr_{0.33}MnO_3$ (ferromagnet) -BaTiO3 (ferroelectric) [35]. Fig. 4(d) shows a rapid increase of the magnetic exchange bias field at the initial stage of electric voltage increment (up to 20 V) and subsequently slowed down at higher electric voltages to achieve the value that falls in the range of magnetic coercivity (362 ± 10 Oe) of the sample.

The magnetic coercivity ($H_C$) has been calculated from the average of the values from positive and negative magnetic field axis of the M(H) loop. The MA50V sample also exhibited similar kind of characters in electric field controlled M(H) loops (Fig. 5(a)), as seen in MA100V sample (Fig. 4). The inset of Fig. 5(a) confirmed a rapid increase of magnetic exchange field ($H_{exb}$) for applied electric field < 20 V and then slowed down to achieve a saturation value for electric voltage above 50 V. The magnetic coercivity ($H_C$) of the sample at higher electric voltages approaches to the value 353 Oe observed at 0 V, whereas MA50V8 sample has achieved the $H_{exb}$ value up to + 370 Oe for electric voltage at + 200 V. The electric field controlled exchange bias field in our system is relatively large in comparison to the values (~ 100 - 220 Oe) observed in hetero-structured multiferroic films that exhibited similar variation of exchange bias shift on increasing the electric voltage [36-37]. As shown in Fig. 5(b), M(H) loop shift of the MA50V8 sample depends on polarity of applied electric voltage and the M(H) loop shifts towards the negative magnetic field axis when measured at - 200 V in comparison to the M(H) loop that is measured in the presence of electric voltage at + 200 V. Subsequently, the magnetic exchange bias shift for negative electric voltages (-10 Oe at -100 V and - 40 Oe at -200 V) is significant with respect to the values obtained at positive voltages + 100 V and + 200 V, respectively. The $H_{exb}(V)$ curve for positive and negative electric field variation is nearly symmetric about a reference line (inset of Fig. 5(b)), which lies in between $H_{exb}(+V)$ and $H_{exb}(-V)$ curves, and of course, not with respect to $H_C(0\ V)$ line.

We have further examined the electric field controlled magnetic state of the samples (Fig. 6 for MA100V8 sample and Fig. 7 for MA50V8 sample) at room temperature through time dependence of magnetization (in-field magnetic relaxation) measurement at constant magnetic field 5 kOe. The in-field magnetic relaxation in both of the Ga doped samples is more or less in similar character (i.e., M(t) decreases in the presence of 5 kOe field) as in hematite sample (Fig. 3(b)). Fig. 6(a) shows the example of magnetic relaxation of MA100V sample in the presence of 5 kOe and at 0 V. The magnetization in both the samples is highly sensitive and switchable under ON-OFF mode and reversal of polarity of the applied electric voltage, as measured at + 100 V (Fig. 6(b), Fig. 7(a)), at -100 V (Fig. 6(c), Fig. 7(b)), at cyclic order of polarity change with sequence 0 V→+ 100 V→ 0 V → -100 V (Fig. 6(d), Fig. 7(c)). As shown in Fig. 7 (d-e), the magnetization in both the samples showed a sudden jump in response to the change of applied voltage either from ON to OFF or OFF to ON modes. A complete reversal of the sign change of magnetization vectors (negative to positive or vice versa) may not be observed upon reversing the polarity of applied electric voltage during in-field magnetic relaxation of spin moments (coloured symbol), but the material instantly jumps to higher magnetic state at the time of applying positive voltage ((ΔM nearly 1.36 %) and jumps to lower magnetic state at the time of applying negative voltage (magnetization change ~ 1.14 %) with respect to the magnetization state at 0 V. The in-field magnetization being in the meta-stable state relaxed slowly with time even in the presence of electric field, irrespective of positive or negative sign, towards achieving the magnetization state at 0 V. However, there exists a gap (ΔM nearly 0.13-0.26 %) between the relaxed magnetic state just before and after switching the ON/OFF modes of electric voltage. It indicates that switched magnetic state even after relaxation with time is different from the relaxed state at 0 V. The existence of magnetization gap during M(t) measurement is irrespective of the magnitude, polarity and cycling of electric voltage. The relaxation of magnetization under simultaneous presence of electric field and magnetic field may be related to the kinetic energy transfer of charge-spin carriers. Fig. 6 (e-f) showed that the relaxation rate is relatively fast at higher electric voltage (at ± 200 V), where the in-field magnetization under 5 kOe relaxed rapidly to the magnetization state either at 0 V preceding to the application of electric voltage or even further relaxed. Such rapid relaxation of magnetization under high electric field may be affected by the leakage of polarization that cannot retain the switched magnetization state for long time. Fig. 7(f) shows the in-field magnetic relaxation where the applied electric voltage is increased step-wise (size 50 V) up to + 500 V for every 50 s interval during measurement time up to 600 s. This experiment confirms a systematic increase of magnetization (magnetic spin order) with the increase of positive electric voltage. As shown in Fig. 7(g), the change of magnetization ($\Delta M$ (%) = $\frac{(M(V)-M(0))*100}{M(0)}$) of the base and peak values of M(t) data under electric voltage continuously increased with the increase of electric voltage in comparison to the M(t) data at 0 V (during first 50 s of measurement). The change of magnetization is found to be 2.51 % and 1.23 % for peak and base values in the presence of electric voltage 500 V during last 50 s of measurement time. The gap between $\Delta M$ (%) at peak and base line increases with applied voltage, most probably due to relaxation of induced magnetization associated with polarization at higher voltage.

## 4. Discussion

We observed that the parent system hematite ($\alpha$-$Fe_2O_3$) does not show any signature of ferroelectric properties and electric field induced magnetic exchange bias effect. On the other hand, Ga doped hematite system exhibited enhanced ferromagnetic properties, as well as ferroelectric and magneto-electric (multiferroelectric) properties. The Ga doped hematite is a new addition in the family of non-traditional multiferroics, which are classified into type-I (e.g. $BiFeO_3$ where ferroelectricity and magnetism have different origins) and type-II (e.g., $TbMnO_3$ where ferroelectric order is induced by magnetic spin order) [38]. Detailed works in future may properly fix the class of Ga doped hematite system, whether type-I or type-II multiferroics. Our work suggests that Ga doped hematite could be of type-II multiferroic, where electric polarization may be small (~1-0.01 $\mu C/cm^2$). Most of the reported materials, either naturally existed or artificially designed, indicate multiferroic properties mainly at low temperatures and unsuitable for room temperature device applications. As of date, $BiFeO_3$ is the reliable single-phased material that exhibited multiferroic properties at room temperature. However, large strain effect determined the exhibition of considerably large value of electric polarization up to 130 µC/cm$^2$ and electric field controlled magnetic state in thin film form of BiFeO$_3$ [39-42]. There are many similarities in the crystal structure and magnetic properties of BiFeO$_3$ and Ga doped hematite. Both BiFeO$_3$ and Ga doped hematite stabilized in rhombohedral structure with R$\bar{3}$C space group. The system is magnetically layered spins structure, where in-plane Fe$^{3+}$ spins ordered ferromagnetically and Fe$^{3+}$ spins in alternating planes (A and B) along off-plane direction ordered antiferromagnetically by superexchange interactions (Fe$^{3+}_A$-O$^{2-}$-Fe$^{3+}_B$). The weak ferromagnetism arises in both the systems due to canting of antiferromagnetically aligned spins controlled by Dzyaloshinskii–Moriya (DM) interaction [$\sim \vec{D}\cdot(\vec{S_n} \times \vec{S_{n+1}})$]. In Ga doped hematite system a dielectric peak, indicating magneto-electric effect [27-28], appeared in the temperature regime (290 K-310 K) where Fe$^{3+}$ spins started to flip from in-plane direction (canted FM state) to out of plane direction with up spins for planes A and down spins for planes B, respectively, and resulted in AFM order of Fe$^{3+}$ spins below Morin transition (~ 260 K). This is the indication that variation in canted spin order has a strong effect on magneto-electric coupling in Ga doped hematite. We try to understand below the mechanisms of magneto-electric properties in Ga doped hematite using existing theoretical models and experimental work on different systems.

Various models have been proposed for the spin induced ferroelectricity. In material, like Ga doped hematite, electronic charge carriers with ½ spin hop between two cations via oxygen anions, i.e., Fe$^{3\pm\delta}$-O$^{2-}$-Fe$^{3\mp\delta}$ superexchange paths [43]. In addition to the current component due to electronic charge carriers, spin current component has played a significant role on determining the magneto-transport properties in Ga doped hematite system [29]. In such case, the spin current flowing between two canted spin sites S$_i$ and S$_j$ is non-zero when $\vec{S}_i \times \vec{S}_j \neq 0$. It is readily available along off-plane direction or in in-plane direction if there is a spiral kind of distribution of spin canting in the rhombohedral structure of Ga doped hematite system. Inspite of the lack of large non-centrosymmetric displacement of cations, as in BaTiO$_3$, the antisymmetric spin exchange interactions $\hat{e}_{ij} \times (\vec{S}_i \times \vec{S}_j)$, where $\hat{e}_{ij}$ is the unit vector connecting spins at sites i and j, can produce macroscopically observable polarization ($\vec{P}_{ij}$) under the influence of relativistic spin–orbit coupling [44]. This polarization can be of genuine electric origin, but its magnitude depends on the displacement of intervening ligand ions ($O^{-2}$) that favour the DM interaction. In order to get an appreciable polarization, the net induced striction should be non-zero and spin modulation should be commensurate with crystal lattice. Such non-zero striction depends on the breaking of inversion symmetry in the spin order along alternating A and B planes of rhombohedral structure. The magnetically tunable ferroelectricity based on the exchange striction mechanism is observed in a family of orthoferrites RFeO$_3$ (R = Gd, Tb, Dy). The low temperature multiferroic properties (electric field controlled magnetism) in RFeO$_3$ type perovskites have no use for room temperature electro-magnetic devices. In spin induced ferroelectrics, a non-centrosymmetric magnetic spin order breaks the inversion symmetry [13]. Such pre-requisite condition is expected in Ga doped hematite and we understand the response of magnetic spins with magnetic field and electric voltage using a schematic diagram (Fig. 8). One source of breaking the inversion symmetry of spin order is the replacement of Fe atoms in A or B planes by non-magnetic Ga atoms (Fig. 8(a)). The canted spin order in undoped and Ga doped hematite system has been understood from the magnetic properties. The inequivalent exchange strictions between the spin order in A and B planes can enhance magnetization and electric polarization in Ga doped system [26]. There is a possibility that the superexchange path lengths (both short and long) in hematite system [45] can be modified by Ga doping and such structural change can generate electric polarization. We did not observe any significant structural distortion within the limit of X-ray diffraction pattern, but the enhancement of magnon-phonon mode at about 1310 cm$^{-1}$ in Raman spectra indicated the enhancement of spin-lattice coupling in Ga doped hematite samples [26]. Second source of breaking the inversion symmetry of spin order is the magnetically inequivalent core-shell spin structure in nano-sized grains of Ga doped hematite system (Fig. 8(b)). In a typical AFM system with layered spin structure, the net magnetic moment (<μ>) between two adjacent layers A and B is zero and it is non-zero in the case of canted FM or AFM (Fig. 8(c). In spin canted system, the orientation of net magnetic moment vector in the presence of external magnetic field ($H_{ext}$) with respect to local easy axis (EA) is determined by the resultant free energy E = $E_{FM}$+ $E_{AFM}$ + $E_{coupling}$, where $E_{FM}$, $E_{AFM}$, and $E_{coupling}$ represent the free energy terms for FM layer, AFM layer and interlayer coupling, respectively [46]. When the field corresponds to AFM interactions ($H_{AFM}$) dominates and greater than $H_{ext}$, the magnetization of the system in the presence of external magnetic field, as we observe in the presence of 5 kOe, can relax with time (in-field magnetic relaxation) to achieve a magnetic state near to bulk AFM (core part of the grains); rather than showing an increase of magnetization with time as in the case of a disordered FM [34]. In the materials like BiFeO$_3$, which is structurally and magnetically similar to Ga doped hematite system, a direct 180$^0$ reversal of DM vector by ferroelectric polarization is forbidden from symmetry point of view of the thermodynamic ground state. However, first-principle calculations [41] predicted a deterministic reversal of the DM vector and canted spin moments using an electric field at room temperature. The coupling between DM vector and polarization can switch the magnetization by 180 $^0$. A complete reversal of magnetization (180 $^0$ rotation) driven by electric field has been predicted in a perpendicularly magnetized nanomagnetic thin film [47]. The electric field response of local magnetization vector M(x, t) depends on the spatial distribution of magnetization vector components inside the magnetic domains and on the temporal evolution of the magnetization vector, determined by Landau-Lifshitz-Gilbert equation: $\frac{\partial \vec{M}}{\partial t} = -\gamma_0 (\vec{M} \times \vec{H}_{eff}) + \frac{\alpha}{M_S}(\vec{M} \times \frac{\partial \vec{M}}{\partial t})$, where $\gamma_0$ and α are the gyromagnetic ratio and the Gilbert damping factor. A competition between the spin torque (first term) and damping torque (second term) determines the orientation of the magnetization vectors either in one side of the interfacial plane or complete reversal about the plane. The I(V) and I(t) measurements under magnetic field ON-OFF modes confirm that the electrical charge carriers in our material experience a torque (as in Landau-Lifshitz-Gilbert equation) or scattering effect on applying the magnetic field (due to Lorentz force) during I-V measurement. The effect of a sudden torque on charge carriers is clearly distinguished while the magnetic field is made OFF from its ON state, resulting in a decrease of effective current flow through the sample under dc electric bias voltage. On the other hand, while electrical current flows through the sample, the sweeping (increasing) of electric field not only increased the dc current, but also generates some magnetic field whose effect may not be much different from the application of external magnetic field (ON mode) with moderate strength. Hence, the electric field controlled moving charge carriers may not sense a separate identity of the external magnetic field while electric voltage is in increasing mode, unlike a sudden torque experienced during the OFF mode of magnetic field. If we look carefully the magnetization switching in the M(t) and M(H) curves of our material, it may be understood that all the spins in the magnetic domains are not participating in the reversal process under electric field. A fraction (roughly 1 %) of the total spins in a magnetic domain, most probably at the domain walls or shell part of the grains (Fig. 8(e-f)), reverses their magnetic spin directions ($180^0$ rotation) upon reversal the polarity of applied electric voltage and rest of the spins (interior to the domain) are reluctantly participating in the reversal process. The above argument is well consistent to the fact that a completely $180^0$ reversal of the magnetization vectors by reversing the electric field polarity is macroscopically observed in Ga doped hematite system during M(t) measurement, of course with reference to the magnetization state at 0 V.

On the other hand, the system showed large electric field controlled exchange bias effect. Some materials, either naturally existed (β-NaFeO$_2$ [48]; M-type hexaferrite BaFe$_{10.2}$Sc$_{1.8}$O$_{19}$ [49]) or synthetic ((LuFeO$_3$)$_9$/(LuFe$_2$O$_4$)$_1$ super lattice [50], nanosheets of Ti$_{0.8}$Co$_{0.2}$O$_2$/Ca$_2$Nb$_3$O$_{10}$/Ti$_{0.8}$Co$_{0.2}$O$_2$ superlattices [51]) or theoretically predicted (R$_2$NiMnO$_6$/La$_2$NiMnO$_6$ [39]), exhibited electric field controlled magnetic state and exchange bias near to room temperature. Most of the engineered magneto-electrics are bi-layered/multilayered structure of FM/AFM films on ferroelectric material. Several mechanisms are involved at the interfaces of ferroelectric and ferromagnetic films [52], e.g., (i) electric field dependent magnetic spin order at domain walls, (ii) spin–orbit interaction of electrons, (iii) strain mediated coupling, and (iv) exchange bias. The electric field controlled exchange bias was first explored at the heterostructured interface of Cr$_2$O$_3$ (Co/Pt)$_3$ by electric field controlled cooling through T$_N$ of Cr$_2$O$_3$ [37]. The observed electric field controlled magnetism in these heterostructured systems, where magnetization state changed with the polarity of applied voltage, was attributed to the strain-mediated magnetoelectric coupling at the heterojunction of ferromagnet and ferroelectric [35]. The present Ga doped hematite is not in thin film form and excluded from strain mediated contribution for exhibiting magnetoelectric coupling. Hence, the observed magnetoelectric properties are considered to be intrinsic of the material. Magnetic state of a system is defined by the magnetic exchange interactions. The change of magnetic state by electric field indicates the spin-orbit coupling or equivalence additional magnetic field in the system. One important aspect may be unnoticed to many researchers that layer kind lattice structure, whether single phased (BiFeO$_3$) or multi-phased (La$_{0.67}$Sr$_{0.33}$MnO$_3$/BaTiO$_3$) materials, seems to be more suitable for exhibiting multiferroic properties at room temperature. The present system also intrinsically forms layered structure of alternative AFM spins along the direction perpendicular to the spin layers. Skumryev et al. [36] demonstrated the magnetization reversal and magnetic exchange bias (EB) of a FM

Ni81Fe19 (Py) film deposited on AFM multiferroic (LuMnO3) single crystal as an effect of electric-field driven decoupling between ferroelectric and antiferromagnetic domain walls. They proposed clamped and unclamped AFM domain walls (AF-DWs) at the interfaces with FE domains. A coupling between clamped AFM-DWs and FE domains exerts electric field controlled torque on FM moments, where as unclamped AF-DWs do not play significant role in electric field controlled magnetization switching due to lack of coupling with FE domains. In case of hematite system, we observed electric field controlled magnetization switching, but no magnetic exchange bias effect. We understand that a unique coupling between magnetic spin order and ferroelectric polarization at the domain walls (DWs) [35-36, 53] or core-shell interface plays a dominant role in exhibiting magnetic exchange bias effect in Ga doped hematite system. In case of M(H) measurement, the core part of the spin structure also contributes in magnetic domain wall motion/domain rotation process. This results in an irreversible shift of the M(H) loop in the presence of electric field, unlike the electric field controlled $180^0$ reversal of magnetization spins mainly at domain walls/shells during M(t) process.

## 5. Conclusions

We conclude that doping of non-magnetic Ga in rhombohedral structure of α-$Fe_2O_3$ enhances multiferro-electric order. The I-V characteristics under magnetic field provided useful information of the magneto-electric effect in the sample. The change in the I-V curve is associated with the dynamics of charge carriers, in addition to the magnetic field effect. On the other hand, ferromagnetic loop, being a universal feature irrespective of metal, semiconductor or insulator, under external electric field was used to detect the existence of magneto-electric coupling in Ga doped hematite system, which exhibited semiconductor behavior and it may belong to the class of semiconductor-magnetoelectric material. The Ga doped hematite system exhibited similar lattice (rhombohedral) and magnetic (spin canting is the cause of net ferromagnetic moment) structure as in $BiFeO_3$. The present system is also a lead free system like $BiFeO_3$. The electric polarization in Ga doped hematite system (~0.1 $\mu C/cm^2$) may be small in comparison to $BiFeO_3$ film (~ 10-130 $\mu C/cm^2$) that is largely affected by strain effect, but the magneto-electric coupling in Ga doped hematite system is free from strain mediated coupling effect and considered to be intrinsic. Most significantly, electric field controlled magnetic state in Ga doped hematite system has a tremendous impact on developing spintronics based new technology. The magnetic exchange bias is generally produced after high magnetic field cooling the magnetically bi-layered (FM/AFM)/exchange-coupled systems at low temperatures (<100 K) from higher temperature. The magnetic exchange bias effect in Ga doped hematite system at room temperature without any magnetic field cooling process and controlled by external electric field alone is enough important for non-volatile memory and electro-magnetic switching devices. The sensing properties of the material for simultaneously presence and change of the polarity of electric and magnetic fields can be used for developing electro-magnetic sensors.

**Acknowledgment**

Authors thank to CIF, Pondicherry University for providing some experimental facilities. RNB acknowledges research grants from Department of Science and Technology (NO. SR/S2/CMP-0025/2011) and Council of Scientific and Industrial Research (No. 03(1222)/12/ EMR_II), Government of India for carrying out this experimental work.

**Figure captions**

Fig. 1 P-E loops for MA50V sample (left side) measured at different electric voltage (a), frequency (b), magnetic field (c), and corresponding variation of $E_C$ and $P_R$ in right side.

Fig. 2 (a) Current -voltage characteristics of the MA100V sample at magnetic field zero and at 15 kOe ON-OFF mode (shown by vertical lines), (b) first order derivative of the I-V curves.

Fig. 3 (a) M(H) loops of hematite sample measured at different electric voltages, magnetic relaxation at 5 kOe for 0 V (b), for ON-OFF modes of +100 V (c), for ON-OFF modes of -100 V.

Fig. 4(a) Room temperature M(H) loops at different measurement voltages, (b) magnified M(H) loops, (c) M(H) loops at 0 V and ± 200 V for MA100V sample, (d) variation of exchange bias with different positive applied voltages.

Fig. 5 Magnified form of room temperature M(H) loops of MA50V sample, measured at different electric voltages (a), shown for +200 V and -200 V with respect to 0 V loop (b). Insets show the variation of magnetic exchange bias field for different electric voltage (a) and for electric bias voltages at ±100 V and ± 200 V (b).

Fig. 6 Room temperature time dependent magnetic moment at a magnetic field of 5 kOe and different voltage conditions (a) 0V, (b) 0V and 100V, (c) 0V and +100V, and (d) 0V and -100V, (e) 0V and +200V, (f) 0V and -200V.

Fig. 7 Time dependent magnetic moment at a magnetic field of 5 kOe and voltage conditions (a) 0V and +100V, (b) 0V and -100V, (c) 0V and 100 V, (d) branch of (a), (e) branch of (b), (f) magnetization at voltage 0-500 V, (g) change of magnetization.

Fig. 8 A schematic diagram of the spin order between two planes in α-Fe$_2$O$_3$ and Ga doped α-Fe$_2$O$_3$ (a), Core-shell spin structure in a grain of Ga doped α-Fe$_2$O$_3$ (b), in-field magnetic relaxation at V = 0 (c) and in the presence of constant voltage (d), response of spin vectors during M(H) measurement in the presence of constant +ve (e) and −ve (f) voltage.

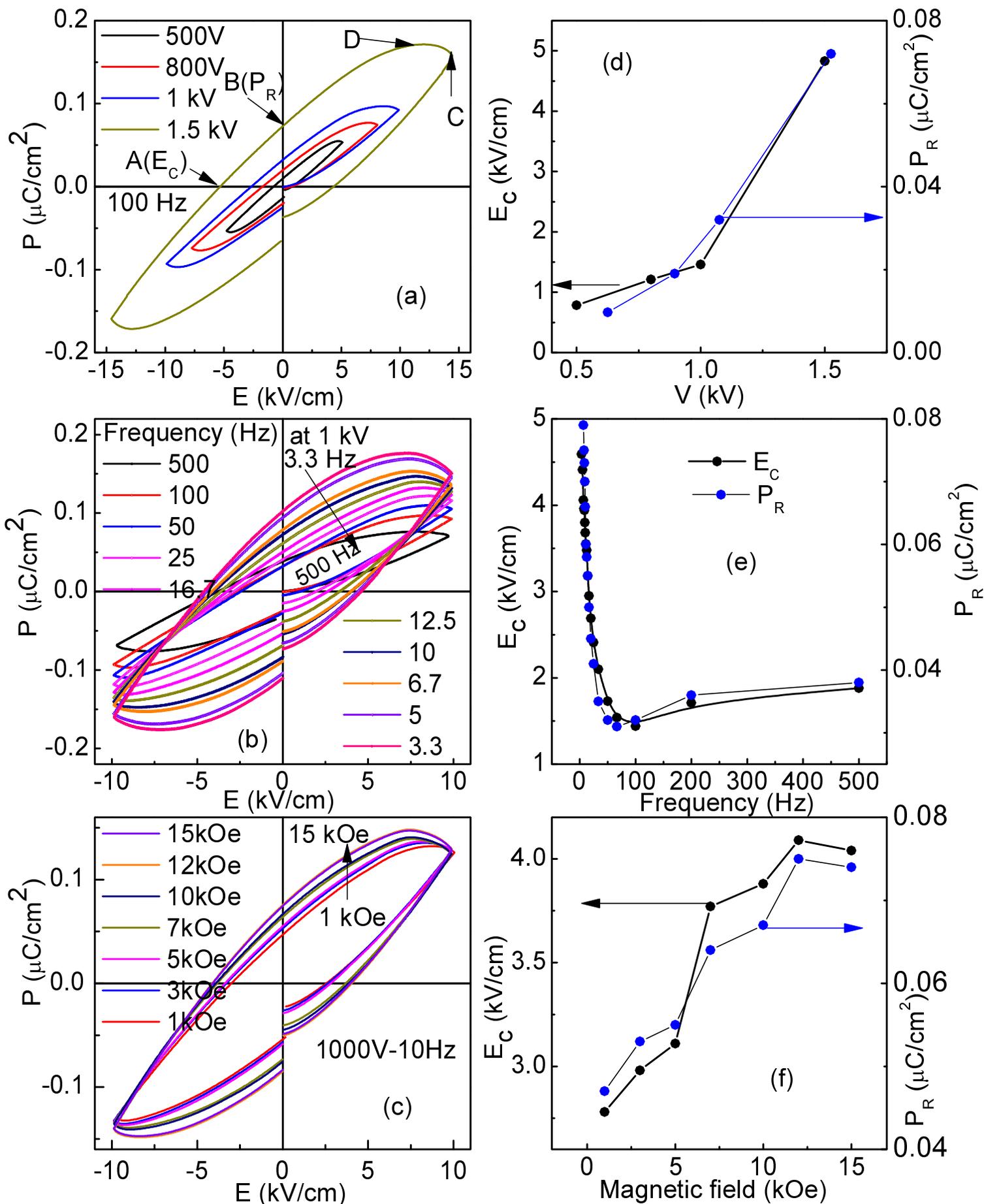

Fig. 1 P-E loops for MA50V sample (left side) measured at different electric voltage (a), frequency (b), magnetic field (c), and corresponding variation of $E_C$ and $P_R$ in right side.

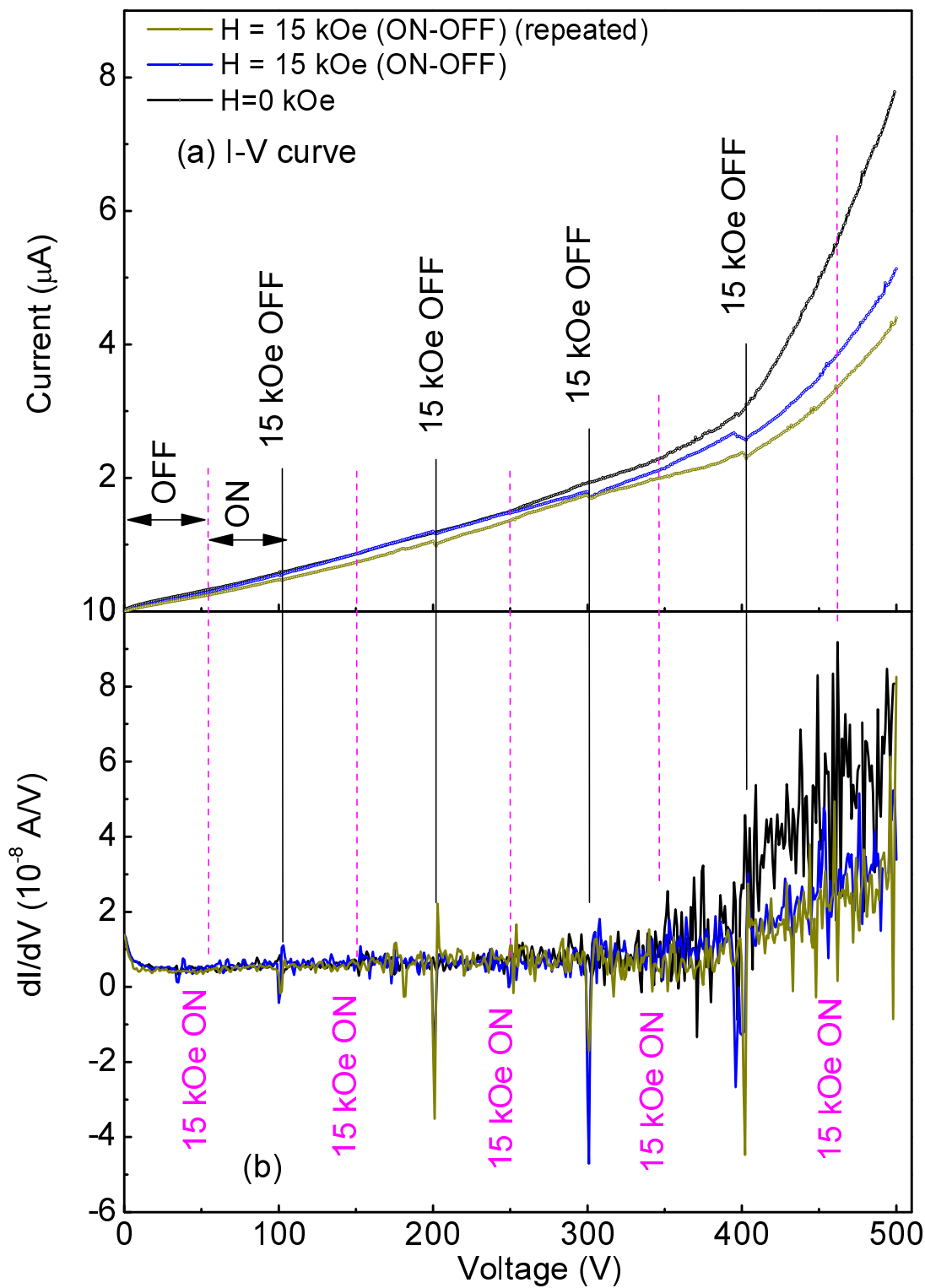

Fig. 2 (a) Current-voltage characteristics of the MA100V sample at magnetic field zero and at 15 kOe ON-OFF mode (shown by vertical lines), (b) first order derivative of the I-V curves

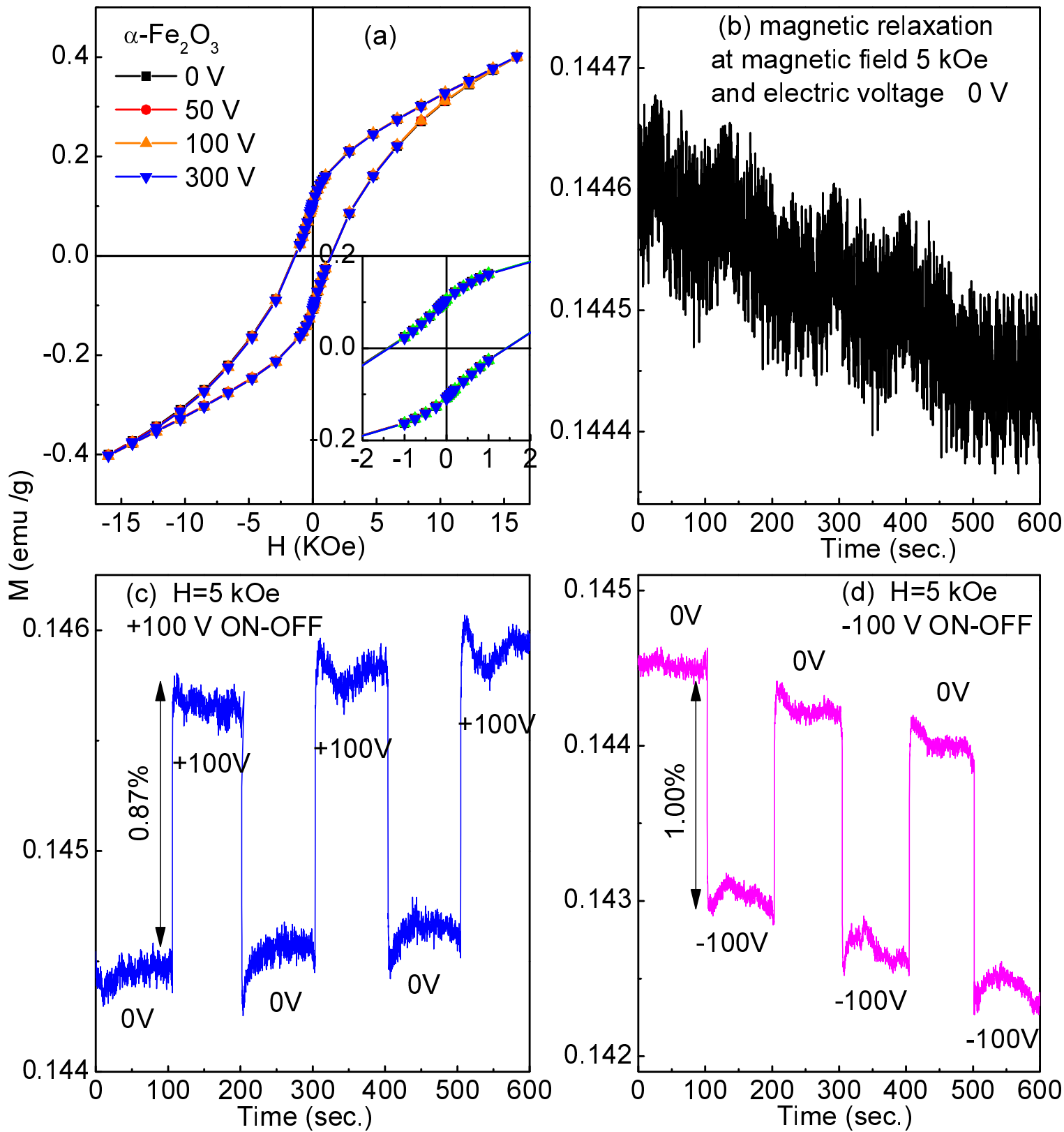

Fig. 3 (a) M(H) loops of hematite sample measured at different electric voltages, magnetic relaxation at 5 kOe for 0 V (b), for ON-OFF modes of +100 V (c), for ON-OFI modes of -100 V.

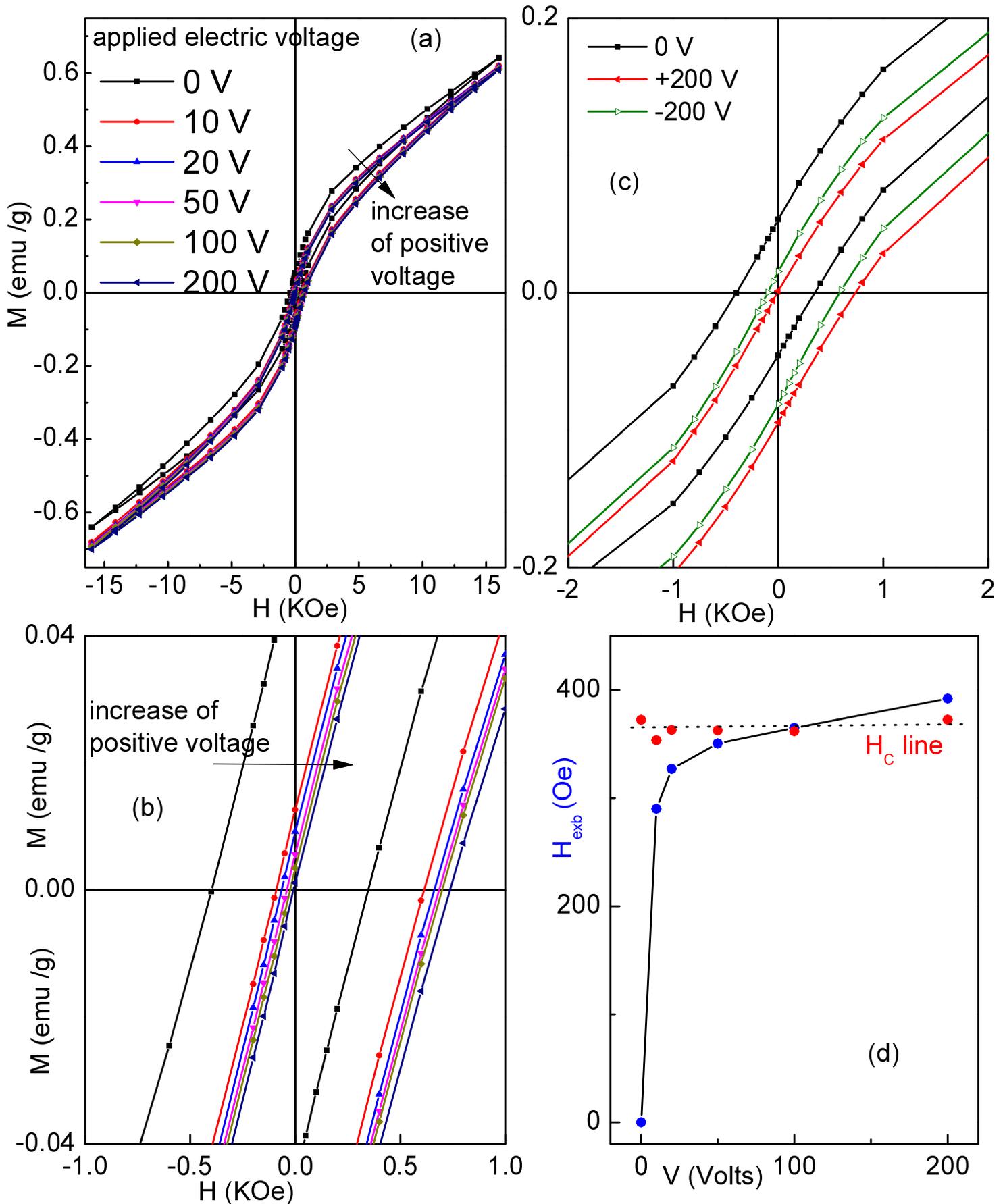

Fig. 4(a) Room temperature M(H) loops at different measurement voltages, (b) magnified M(H) loops, (c) M(H) loops at 0 V and ± 200 V for MA100V sample, (d) variation of exchange bias with different positive applied voltages.

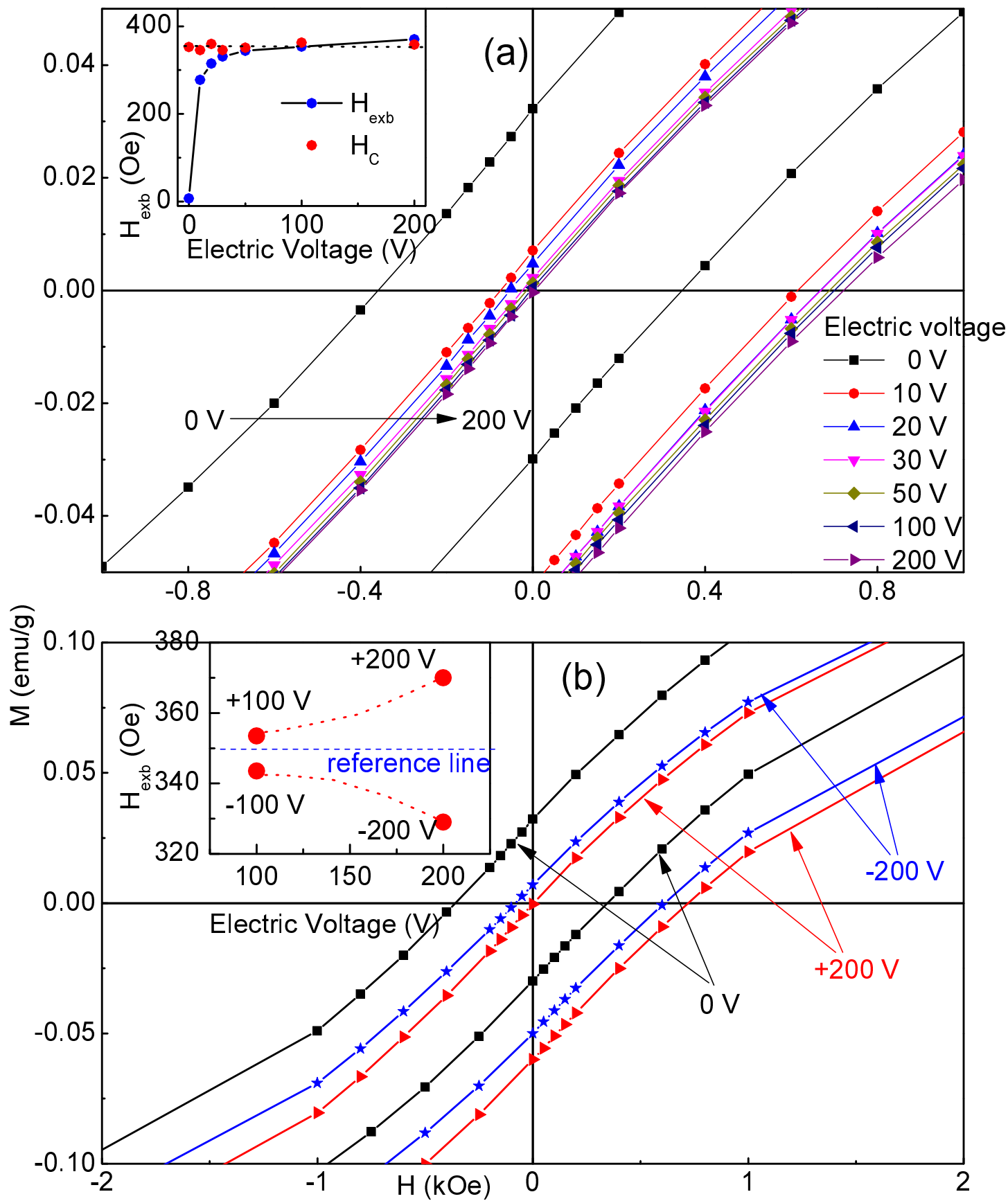

Fig. 5 Magnified form of room temperature M(H) loops of MA50V sample, measured at different electric voltages (a), shown for +200 V and -200 V with respect to 0 V loop (b). Insets show the variation of magnetic exchange bias field for different electric voltage (a) and for electric bias voltages at ± 100 V and ± 200 V (b).

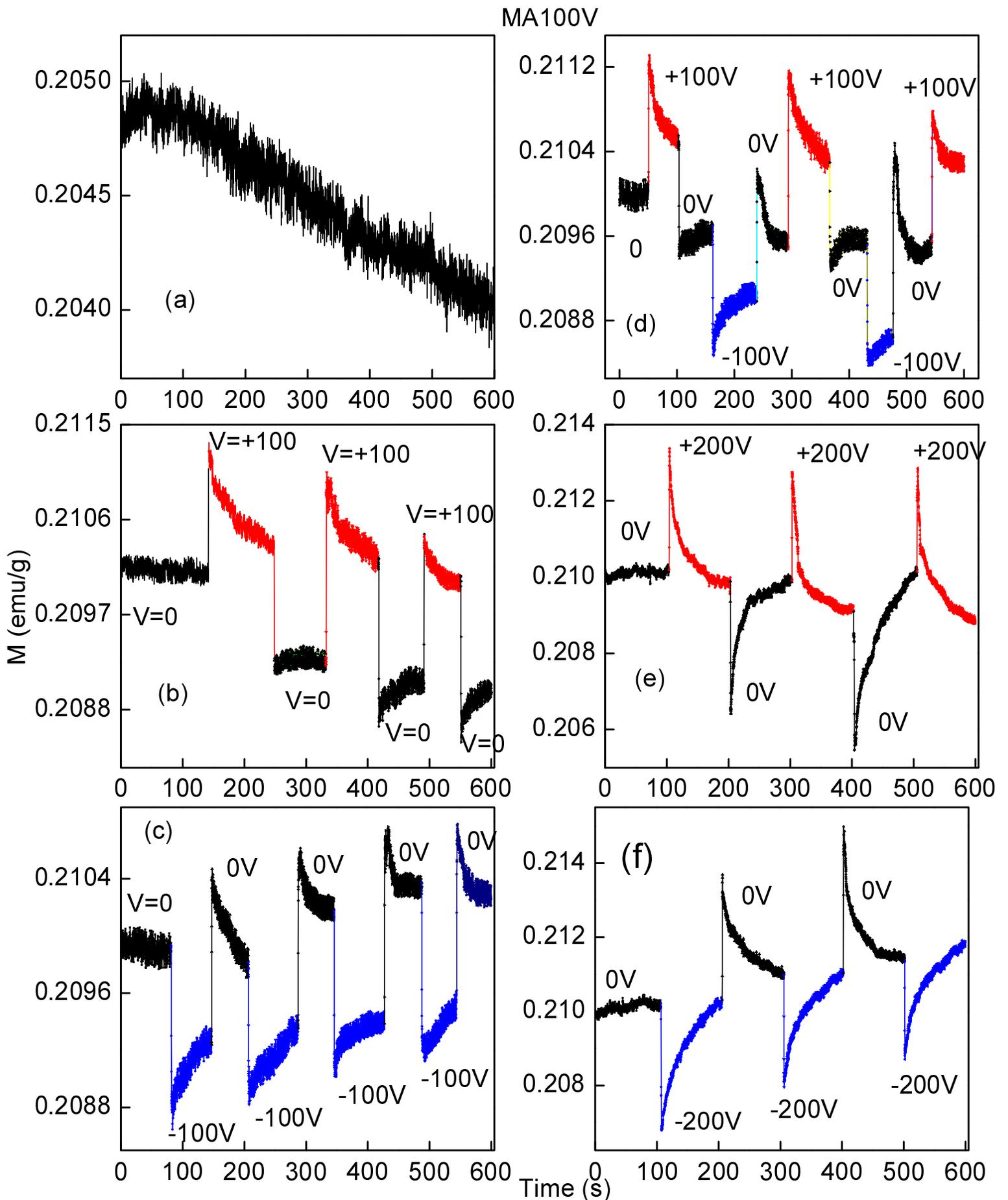

Fig. 6 Room temperature time dependent magnetic moment at a magnetic field of 5 kOe and different voltage conditions (a) 0V, (b) 0V and ±100V, (c) 0V and +100V, and (d) 0V and -100V, (e) 0V and +200V, (f) 0V and -200V.

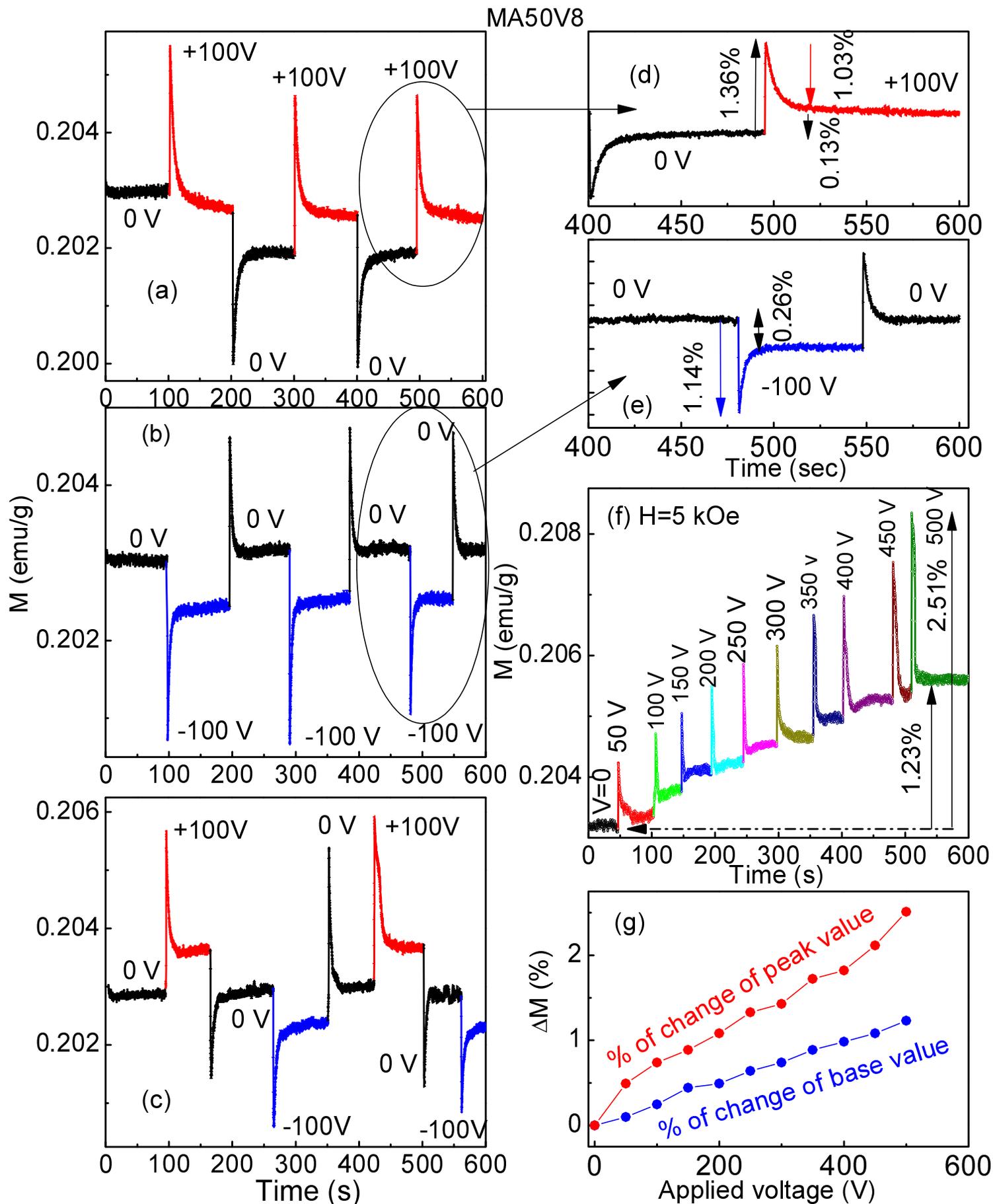

Fig. 7 Time dependent magnetic moment at a magnetic field of 5 kOe and voltage conditions (a) 0V and +100V, (b) 0V and −100V, (c) 0V and ±100 V, (d) branch of (a), (e) branch of (b), (f) magnetization at voltage 0-500 V, (g) change of magnetization.

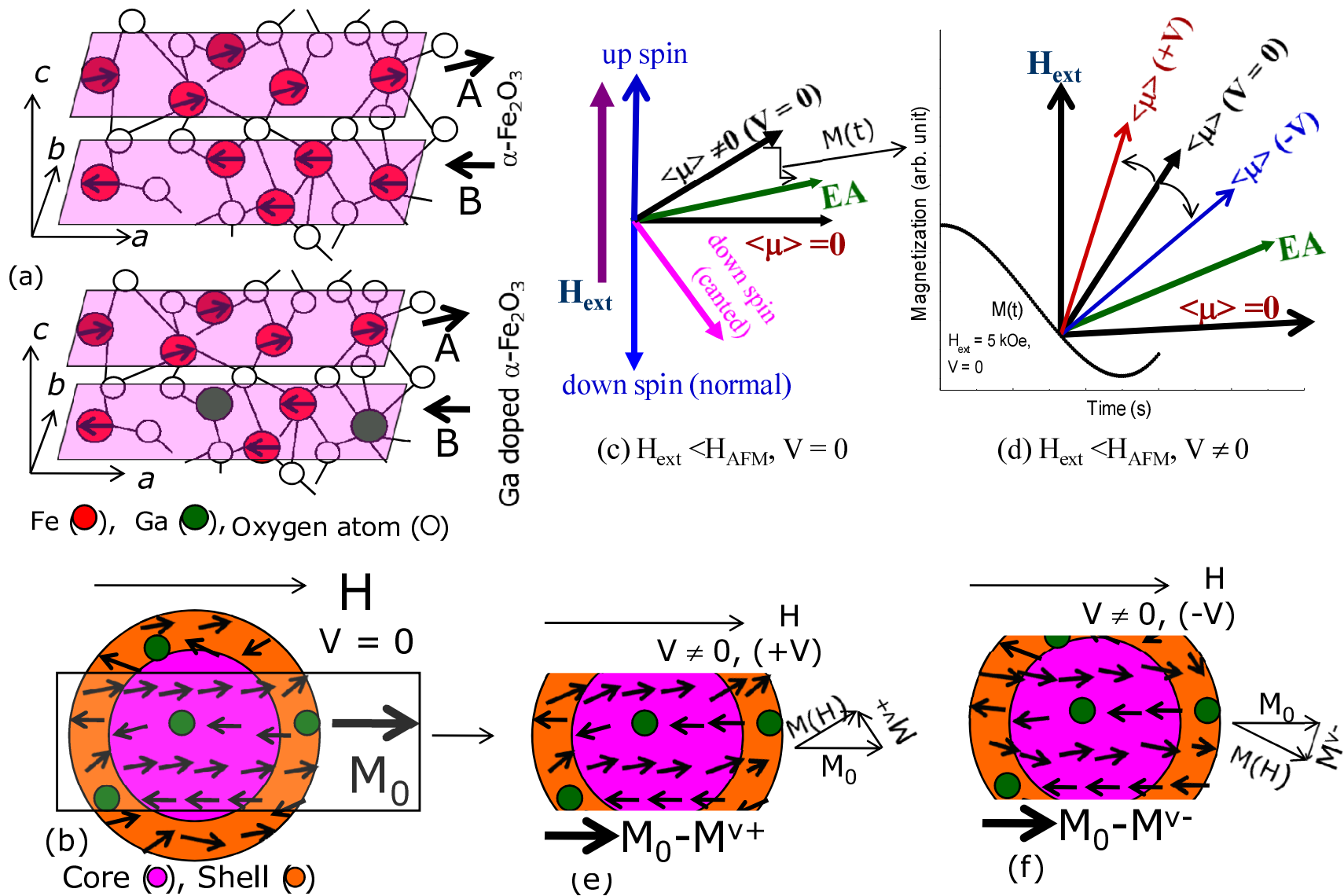

Fig. 8 A schematic diagram of the spin order between two planes in α-Fe$_2$O$_3$ and Ga doped α-Fe$_2$O$_3$ (a), Core-shell spin structure in a grain of Ga doped α-Fe$_2$O$_3$ (b), in-field magnetic relaxation at V = 0 (c) and in the presence of constant voltage (d), response of spin vectors during M(H) measurement in the presence of constant +ve (e) and –ve (f) voltage.